\documentclass[a4paper,11pt]{article}

\usepackage{pos}
\usepackage{siunitx}
\usepackage{hyperref}

\title{An improved infrastructure for the IceCube realtime system}

\ShortTitle{Improved IceCube realtime infrastructure}

\author{The IceCube Collaboration \\{\normalsize \normalfont(a complete list of authors can be found at the end of the proceedings)}\\}

\emailAdd{lincetto@astro.ruhr-uni-bochum.de}

\abstract{

The IceCube realtime alert system has been operating since 2016. It provides prompt alerts on high-energy neutrino events to the astroparticle physics community. The localization regions for the incoming direction of neutrinos are published through NASA's  Gamma-ray Coordinate Network (GCN). The IceCube realtime system consists of infrastructure dedicated to the selection of alert events, the reconstruction of their topology and arrival direction, the calculation of directional uncertainty contours and the distribution of the event information through public alert networks. Using a message-based workflow management system, a dedicated software (SkyDriver) provides a representational state transfer (REST) interface to parallelized reconstruction algorithms. In this contribution, we outline the improvements of the internal infrastructure of the IceCube realtime system that aims to streamline the internal handling of neutrino events, their distribution to the SkyDriver interface, the collection of the reconstruction results as well as their conversion into human- and machine-readable alerts to be publicly distributed through different alert networks. An approach for the long-term storage and cataloging of alert events according to findability, accessibility, interoperability and reusability (FAIR) principles is outlined.

\vspace{4mm}
{\bfseries Corresponding authors:}

Eric Evans-Jacquez$^{1}$, Massimiliano Lincetto$^{2*}$, Benedikt Riedel$^{1}$, David Schultz$^{1}$, Tianlu Yuan$^{1}$\\
{$^{1}$ \itshape Dept. of Physics and Wisconsin IceCube Particle Astrophysics Center, University of Wisconsin-Madison, Madison, WI 53706, USA}\\
{$^{2}$ \itshape Fakultät für Physik \& Astronomie, Ruhr-Universität Bochum, D-44780 Bochum, Germany}\\[4mm]
$^*$ Presenter

\ConferenceLogo{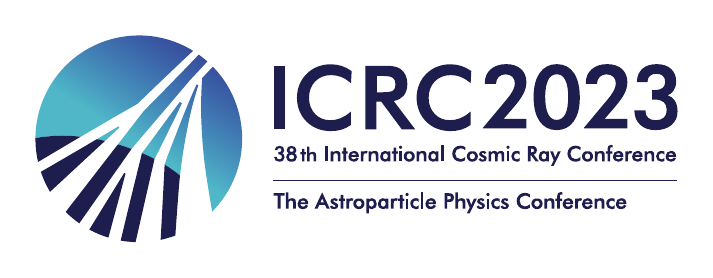}

\FullConference{The 38th International Cosmic Ray Conference (ICRC2023)\\ 26 July -- 3 August, 2023\\ Nagoya, Japan}
}

\begin{document}

\maketitle


\section{Introduction}
\label{s:intro}

The IceCube realtime alert system has been established in 2016 to provide the astronomy and astroparticle community with timely alerts on the detection of high-energy neutrinos that have a moderate-to-high probability of being astrophysical. Since 2019, the system has been updated with a new set of selection criteria resulting in the implementation of "gold" and "bronze" channels for track alerts. The two provide neutrino alerts with average probability of being astrophysical of 50\% (gold) and 30\% (bronze). For these events, a detailed reconstruction is required to estimate the neutrino arrival direction and its uncertainty. Since this operation is especially computing intensive, it requires the orchestration of a massively parallelized workload on a high-performance-computing (HPC) cluster. The original implementation of this system is based on a custom message distribution protocol and is specifically designed to work with the IceCube in-house HTCondor~\cite{condor-practice} cluster. As such, it cannot be easily adapted to different computing infrastructures. A first porting of the distributed reconstruction system to a commercial cloud computing service has been achieved to process the "IceCat-1" event catalog of alert tracks~\cite{IceCube:2023agq, DVN/SCRUCD_2023}. While effective for batch offline processing, this implementation is not suitable for the purpose of the realtime system. In this work, we outline a redesign of the reconstruction software to improve its portability and scalability. We describe a set of associated improvements to the general alert handling infrastructure of IceCube. Allowing access to a broader pool of computing resources and a higher level of automatization, this will improve the alert response times and reduce the chance of human-induced errors in the reporting of the results. In addition, the new developents aim to improve the current data management practices, pursuing the adherence to the "FAIR" principles for "findability, accessibility, interoperability and reusability" of scientific data.

\section{Current status of real-time alert handling}

The online triggering and filtering system of IceCube processes the data at the South Pole. Candidate neutrino events are subject to preliminary set of reconstructions, to determine their topology, direction and energy. A real-time selection for track-like events (above an energy of \SI{100}{GeV}) produces an event rate of a few mHz dominated by atmospheric backgrounds. This selection was originally introduced in 2008 to search for neutrino multiplets aimed at followups in optical, X-ray, and gamma-ray bands. Such a program is still active under the label of Gamma-ray Follow-Up (GFU)~\cite{IceCube:2023icrc-GFU}.
The same event selection serves as the base for the identification of individual neutrinos with a moderate-to-high probability of having astrophysical origin (candidate alerts). For all track-like events, the summary data obtained from the preliminary set of reconstructions are transmitted to a data center in the Northern hemisphere through a commercial satellite network. For candidate alerts, an additional message carries the full event data recorded by the IceCube digital optical modules, allowing for more sophisticated evaluations and reconstructions to be applied later. The transmission of the data from the Pole and its collection at the North is orchestrated by the IceCube Live (I3Live) control system~\cite{IceCube:2016zyt-INST}. An in-depth description of the IceCube realtime system is given in Ref.~\cite{IceCube:2016cqr-RT}. At the IceCube data center in the North, realtime event data are distributed through a ZeroMQ \cite{ZeroMQ} message queue and stored by I3Live in a private but worldwide-accessible MongoDB\footnote{\url{https://www.mongodb.com}} database. The events received through the message queue are further evaluated. If an event passes the alert selection criteria, a prompt automated alert is issued through the infrastructure of the Astrophysical Multimessenger Observatory Network (AMON)~\cite{AyalaSolares:2019iiy}, to be published and distributed in the form of a machine-readable "notice" on the NASA General Coordinates Network (GCN, formerly Gamma-ray Coordinates Network -- Transient Astronomy Network \cite{NASA-GCN}) platform.

In parallel to the issuance of an alert, the same event is promptly scheduled for a more sophisticated reconstruction exploiting on IceCube computing cluster. The result of the reconstruction is typically obtained in about one to three hours from the issue of the first alert. Upon completion, the results are automatically reported through an internal messaging platform and evaluated by humans. After the status of the detector and the result of the updated reconstruction have been scrutinized, the updated alert information is distributed in the form of a human-readable GCN "circular" and an update of the first machine-readable notice.

\section{The parallelized reconstruction system}

\begin{figure}
    \centering
    \includegraphics[width=0.45\textwidth]{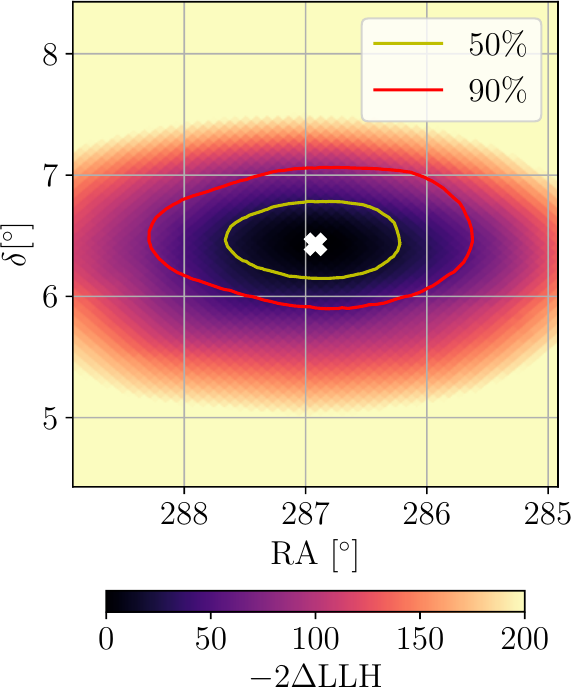}
    \centering
    \caption{Example likelihood map in equatorial coordinates for a neutrino event reconstructed with "Skymap Scanner" from Ref.~\cite{IceCube:2023agq}. The color represents the difference in log-likelihood between the position and the best fit. 50\% and 90\% containment contours are marked with solid yellow and red lines.}
    \label{fig:scan-example}
\end{figure}

To ensure the most accurate reconstruction of a realtime event, the sky area represented in equatorial coordinates (right ascension and declination) is divided into a grid of pixels following the HEALPIX scheme \cite{Gorski:2004by}. The center direction of each pixel is converted to the local detector coordinates (zenith and azimuth), and used as fixed parameter for a maximum-likelihood reconstruction algorithm. The algorithm currently in place for the reconstruction of realtime events \cite{IceCube:2013dkx} estimates best fit parameters for the muon deposited energy and the position of the interaction vertex. Algorithms based on simpler likelihood descriptions may fit for only the latter. The pixel for which the reconstruction yields the best likelihood is taken as the best fit direction for the event. Critical values for the variation of the likelihood around the best fit position are used to determine 50\% and 90\% containment contours for the event localization \cite{IceCube:2023icrc-RTRECO}. We refer to this as the "sky scan" approach. An example of skymap produced by Skymap Scanner is shown in Fig.~\ref{fig:scan-example}. The number of pixels in a HEALPIX grid is defined by the $N_\mathrm{SIDE}$ parameter, $n_{\mathrm{pix}} = 12 N^2_\mathrm{SIDE}$. For a good resolution in the determination of the localization contour, a $N_\mathrm{SIDE}$ value of at least 512 is required. For the full sky, this would require the reconstruction of $\sim \num{3E6}$ pixels. To reduce the number of pixels to test to a sustainable number, the scan is first performed on a coarse pixelization of the sky ($N_\mathrm{SIDE} = 8$). The subset of pixels (typically 12--24) with the best likelihood values is then selected for refinement, and further divided into smaller pixel according to a HEALPIX scheme with an increased value of $N_\mathrm{SIDE}$. Intermediate iterations of the refinement procedure are performed to reach the final resolution around the best fit direction. The current configuration uses an intermediate value of $N_\mathrm{SIDE}$ of 64 and a final value of 512 or 1024. This reconstruction technique is implemented in the "Skymap Scanner" component of the realtime infrastructure.

\begin{figure}
    \centering
    \includegraphics[width=0.75\textwidth]{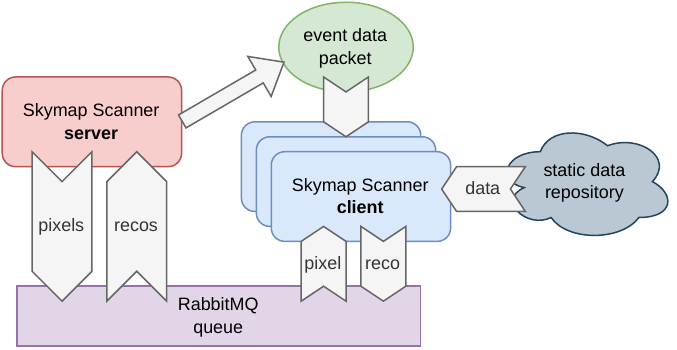}
    \caption{Diagram of the redesigned "Skymap Scanner" component. The scan is initialized by the server, that produces an event data packet to be read by all clients. Each client can further fetch additional required data related to the detector and likelihood descriptions from an external repository. Then, the server and the clients exchange messages about the individual pixels' reconstruction through a RabbitMQ queue.}
    \label{scanner}
\end{figure}

For a given $N_\mathrm{SIDE}$, the reconstruction of each individual pixel is independent of all others. The reconstruction workload can be therefore distributed across an arbitrary number of nodes, each one dedicated to process one or more pixels. In the redesign presented here, the distribution of the pixels to the worker nodes relies on a server-client structure for the "Skymap Scanner" and a RabbitMQ\footnote{\url{https://www.rabbitmq.com}} publish/subscribe message queue~\cite{oasis2012advanced} for interprocess communication. Both the server and the client can run under Docker\footnote{\url{https://www.docker.com}} or Apptainer\footnote{\url{https://apptainer.org}} (formerly Singularity) containers. The client application running on a worker node relies on a set of static data. These include the tables describing the likelihood functions used by reconstruction method~\cite{IceCube:2023icrc-SPLINE} plus the geometry, calibration and detector status information~\cite{IceCube:2016zyt-INST} for the data taking period of interest. A data staging system is implemented: the required static data are either shipped with the container, read from the distributed CernVM-File System (CVMFS) \cite{Buncic_2010-CVMFS}, or automatically fetched from an online file server at runtime. When the application is started, the clients are initialized by reading a "startup" data packet created by the server. This includes the neutrino event data and the coordinates to access a shared RabbitMQ message queue. After both server and clients have started, the essential information about the pixels to reconstruct and the result of said reconstruction are exchanged through the RabbitMQ infrastructure. An outline of this design is shown in Fig.~\ref{scanner}

Thanks to the newly adopted technologies, the achieved design is highly scalable and can be easily run on different computing infrastructures. While the original "Skymap Scanner" required access to the IceCube dedicated HTCondor cluster, the current version can benefit from the broader resource pool of OpenScienceGrid \cite{OSG}. Ultimately, it will be possible to run the realtime reconstruction on commercial cloud computing services. For this type of workload, such resource allocation is far more efficient and sustainable if compared to traditional scientific HPC infrastructures that are not well suited for tasks requiring a large number of nodes for short times.

Furthermore, the original design was developed around a specific reconstruction algorithm \cite{IceCube:2013dkx}, the redesign provides different reconstructions as individual modules. Each reconstruction module specifies separately the server-side operations, such as the pre-processing of the IceCube DOM data, and the client-side operations that implement the maximum likelihood fit for each pixel. Each reconstruction module can further define a set of runtime configuration parameters in the form of a JavaScript Object Notation (JSON) string\footnote{\url{https://json.org}}. The support for multiple reconstruction methods is instrumental to the planned improvement to the realtime event reconstruction, proposed in \cite{IceCube:2023icrc-RTRECO}.

\section{Orchestration: SkyDriver and REST API}
In order to facilitate the access to the new reconstruction application, we have developed "SkyDriver", a \emph{Software-as-a-Service} (SaaS) solution for neutrino event reconstruction. The role of "SkyDriver" is to automatize the orchestration of the client-server "Skymap Scanner" component, allowing the science user to require event reconstructions and collect their results through a representational state transfer (REST) \cite{fielding2000architectural} application programming interface (API). The REST API allows to control the "SkyDriver" operation by means of HyperText Transfer Protocol (HTTP) requests. In this paradigm, a "POST" request is used to send data to the system while a "GET" request is used to retrieve data from the system. An event reconstruction is initiated with a "POST" request providing the event data (serialized in JSON format) and specifying the required "Skymap Scanner" configuration ($N_\mathrm{SIDE}$ progression, reconstruction algorithm, software version). Upon such request, SkyDriver creates a "manifest" containing the metadata of a reconstruction task, and returns its unique identifier. Through the manifest, SkyDriver provides access to the operation status and progress. Upon completion, the result of the reconstruction can be retrieved with a dedicated "GET" request. Manifests and reconstruction results are permanently stored by "SkyDriver" using an internal database, ensuring the proper persistence of provenance information and long term accessibility and reproducibility of the reconstruction results.

\begin{figure}
    \centering
    \includegraphics[width=0.75\textwidth]{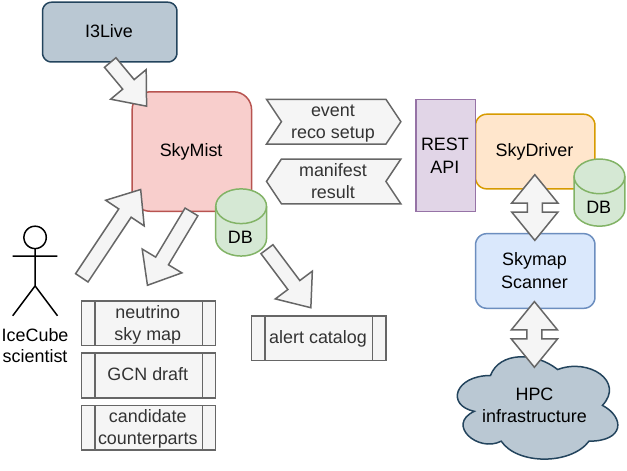}
    \caption{Overview diagram of the improved alert followup infrastructure for IceCube. The "SkyMist" component provides a single point of access to IceCube Live and the event reconstruction services provided by SkyDriver. Most of the data management tasks required by the internal followup of neutrino alerts are automatized.}
    \label{fig:skymist}
\end{figure}

\section{Coding standard and practices}

All the software components of the infrastructure are primarily developed in the Python programming language. The IceCube data processing and reconstruction makes use of a Python interface to C++ code part of the IceCube software framework. The development process is based on "git" for version control, and relies on extensive use of continuous integration, continuous deployment and automated testing. We adopt static type checking\footnote{\url{https://peps.python.org/pep-0484/}} to ensure the robustness and consistency of the codebase. Each reconstruction method implemented in "Skymap Scanner" provides a minimal set of test data obtained through a standard run of the application on a HPC infrastructure. At each update of the code, a test reconstruction is performed on three IceCube test events (two alerts and one simulation), and the result is compared with the static test data. Any error in completing the reconstruction or deviation from the expected result will produce a failed test. The adopted standards ensure the reproducibility of future analyses based on the results of "Skymap Scanner".

\section{A realtime system integration for FAIR and open data}

The redesigned "Skymap Scanner" and "SkyDriver" are integrated in the realtime infrastructure by means of a third component, "SkyMist". "SkyMist" implements the required interfaces to the ZeroMQ realtime queue, the I3Live database and the SkyDriver REST API. "SkyMist" has the primary function of monitoring the stream of realtime data, scheduling the alert reconstruction tasks through "SkyDriver", tracking their progress and reporting the results to the internal IceCube messaging platform. In addition, "SkyMist" allows for the automatic drafting of GCN circulars from the reconstruction results according to pre-defined templates. Although human scrutiny is still required for the sending of circulars as part of the IceCube realtime committee policies, this automatization ensures future consistency in the IceCube neutrino alert communications and removes the risk of transcription errors. IceCube alert GCN circulars routinely report a list of candidate gamma-ray counterparts within the 90\% localization region of the neutrino event. "SkyMist" automatizes the compilation of such list in a standard form, adding an option to report different types of astrophysical transients by querying the public Astro-COLIBRI platform \cite{Reichherzer:2022msg}.

In the last few years, the awareness for good data stewardship practices in the scientific community has been increasing. At the same time, open data policies are becoming an almost ubiquitous requirement for publicly funded scientific projects. The "FAIR" data concept has been introduced to advocate for findability, accessibility, interoperability and reproducibility of scientific data \cite{Wilkinson2016}. With "SkyMist" we aim to implement these principles in the realtime program of IceCube. This is achieved with the automated and centralized storage by "SkyMist" of all the records related to the internal alert handling, reconstruction results and publicly distributed information. For this, a MongoDB instance dedicated to the realtime program is maintained. With this work, we also aim to support the future public data releases of the IceCat alert track catalog, as published through the Harvard Dataverse platform \cite{DVN/SCRUCD_2023}. A diagram of the infrastructure orchestrated by "SkyMist" is shown in Fig.~\ref{fig:skymist}.

\section{Conclusion}

We have redesigned a fundamental component of the IceCube realtime system. The chosen design is modular and scalable, allowing for improved efficiency and sustainability of the computing workloads required by the realtime neutrino event reconstruction. The new system is instrumental to the benchmarking of reconstruction methods aimed at improving the real-time astronomy results of IceCube. By adopting modern coding standards and automated quality control and quality assurance practices, we ensure the long-term reproducibility of the scientific results relying on such reconstruction methods. The ongoing improvement to the IceCube realtime infrastructure will allow for a faster response to alerts, reduced chance of error in the reporting of neutrino information, and the adherence to data findability, accessibility, interoperability and reusability principles (FAIR).

\bibliographystyle{ICRC}
\bibliography{references}

%

\clearpage

\section*{Full Author List: IceCube Collaboration}

\scriptsize
\noindent
R. Abbasi$^{17}$,
M. Ackermann$^{63}$,
J. Adams$^{18}$,
S. K. Agarwalla$^{40,\: 64}$,
J. A. Aguilar$^{12}$,
M. Ahlers$^{22}$,
J.M. Alameddine$^{23}$,
N. M. Amin$^{44}$,
K. Andeen$^{42}$,
G. Anton$^{26}$
% ...
\\
\\
% ...
$^{12}$ Universit{\'e} Libre de Bruxelles, Science Faculty CP230, B-1050 Brussels, Belgium \\
% ...
$^{17}$ Department of Physics, Loyola University Chicago, Chicago, IL 60660, USA \\
$^{18}$ Dept. of Physics and Astronomy, University of Canterbury, Private Bag 4800, Christchurch, New Zealand \\
% ...
$^{22}$ Niels Bohr Institute, University of Copenhagen, DK-2100 Copenhagen, Denmark \\
$^{23}$ Dept. of Physics, TU Dortmund University, D-44221 Dortmund, Germany \\
% ...
$^{26}$ Erlangen Centre for Astroparticle Physics, Friedrich-Alexander-Universit{\"a}t Erlangen-N{\"u}rnberg, D-91058 Erlangen, Germany \\
% ...
$^{40}$ Dept. of Physics and Wisconsin IceCube Particle Astrophysics Center, University of Wisconsin{\textendash}Madison, Madison, WI 53706, USA \\
% ...
$^{42}$ Department of Physics, Marquette University, Milwaukee, WI, 53201, USA \\
% ...
$^{44}$ Bartol Research Institute and Dept. of Physics and Astronomy, University of Delaware, Newark, DE 19716, USA \\
% ...
$^{63}$ Deutsches Elektronen-Synchrotron DESY, Platanenallee 6, 15738 Zeuthen, Germany  \\
$^{64}$ Institute of Physics, Sachivalaya Marg, Sainik School Post, Bhubaneswar 751005, India \\
% ...

\subsection*{Acknowledgements}

\noindent
The authors gratefully acknowledge the support from the following agencies and institutions:
USA {\textendash} U.S. National Science Foundation-Office of Polar Programs,
U.S. National Science Foundation-Physics Division,
U.S. National Science Foundation-EPSCoR,
Wisconsin Alumni Research Foundation,
Center for High Throughput Computing (CHTC) at the University of Wisconsin{\textendash}Madison,
Open Science Grid (OSG),
Advanced Cyberinfrastructure Coordination Ecosystem: Services {\&} Support (ACCESS),
Frontera computing project at the Texas Advanced Computing Center,
U.S. Department of Energy-National Energy Research Scientific Computing Center,
Particle astrophysics research computing center at the University of Maryland,
Institute for Cyber-Enabled Research at Michigan State University,
Astroparticle physics computational facility at Marquette University,
and Cloud credits and support by Google Cloud Platform;
Belgium {\textendash} Funds for Scientific Research (FRS-FNRS and FWO),
FWO Odysseus and Big Science programmes,
and Belgian Federal Science Policy Office (Belspo);
Germany {\textendash} Bundesministerium f{\"u}r Bildung und Forschung (BMBF),
Deutsche Forschungsgemeinschaft (DFG),
Helmholtz Alliance for Astroparticle Physics (HAP),
Initiative and Networking Fund of the Helmholtz Association,
Deutsches Elektronen Synchrotron (DESY),
and High Performance Computing cluster of the RWTH Aachen;
Sweden {\textendash} Swedish Research Council,
Swedish Polar Research Secretariat,
Swedish National Infrastructure for Computing (SNIC),
and Knut and Alice Wallenberg Foundation;
European Union {\textendash} EGI Advanced Computing for research;
Australia {\textendash} Australian Research Council;
Canada {\textendash} Natural Sciences and Engineering Research Council of Canada,
Calcul Qu{\'e}bec, Compute Ontario, Canada Foundation for Innovation, WestGrid, and Compute Canada;
Denmark {\textendash} Villum Fonden, Carlsberg Foundation, and European Commission;
New Zealand {\textendash} Marsden Fund;
Japan {\textendash} Japan Society for Promotion of Science (JSPS)
and Institute for Global Prominent Research (IGPR) of Chiba University;
Korea {\textendash} National Research Foundation of Korea (NRF);
Switzerland {\textendash} Swiss National Science Foundation (SNSF);
United Kingdom {\textendash} Department of Physics, University of Oxford.

\end{document}